          [2004/06/22 1.2 (KOF); 2001/04/25 1.1 (PWD)]

\documentclass[a4paper,twocolumn]{Gaia2004} % European paper size
\usepackage{times}      % for font
\usepackage{epsfig}     % for figure inclusion
\usepackage{natbib}     % for bibliography
\title{Astrometric limits set by surface structure, binarity, microlensing}

\author{U. Bastian, H. Hefele}
\affil{Astronomisches Rechen-Institut, M{\"o}nchhofstr 14, 
69120 Heidelberg, Germany}

\bibpunct{(}{)}{;}{a}{}{,}  % to set bibliography punctuation to A&A style

\begin{document}


\maketitle

\begin{abstract}
We give an assessment of the significance of various known effects which may produce genuine fluctuations of star positions comparable to or larger than Gaia's measurement noise, and which thus may limit the ultimately reachable precision of the determination of parallaxes and proper motions. Stellar granulation is found to be no problem except for a small number of cool supergiants. It will be a serious problem for e.g. Mira star
parallaxes.  Star spots are hard to assess quantitatively. They, too, may be
problematic for supergiants and very cool giants. Binarity is a major problem for the astrometry of bright stars. It will significantly impair the kinematics of stellar aggregates with small velocity dispersion. It will also produce a small proportion of
grossly wrong parallaxes. Gravitational microlensing will not be a problem, although
thousands of highly significant microlensing events will be detected by Gaia. Gravitational macrolensing of quasars will add some noise to their (otherwise ``zero'') proper motions.          

\end{abstract}

\section{Introduction}

Gaia has an extremely broad variety of scientific
goals, see e.g.~ESA 2000 and many papers in the present volume. The central ones, however, are
\begin{itemize}
\item measuring absolute parallaxes for the radius and luminosity
calibration of stars and for the cosmic distance scale.
\item measuring space velocities of many stars for the investigation
of the structure and evolution of our galaxy, the Milky Way.
\end{itemize}
The major limitation of Gaia's ability to do reach these objectives is technical
measurement noise. It is determined by the size of the optics,
the mission duration, the efficiency of Gaia's CCDs, the data
reduction procedures, etc.

But in addition to this purely technical noise there are also
astronomical noise sources, i.e.~genuine fluctuations of 
star positions. There are three major categories of known effects that
may create positional noise: 
\begin{itemize}
\item surface structure of observed objects, including spots and
outer convection zones (granulation) in the case of stars, 
and asymmetric shapes and albedo structures in the case of asteroids.
\item gravitational lensing.
\item stellar multiplicity, including planetary systems.
\end{itemize}

Detecting and measuring such astrometric fluctuations 
constitutes an important part of Gaia's scientific motivation
and outputs: Binary motion measured by Gaia is the most powerful
foreseeable method to find a big unbiased sample of extrasolar
giant gas planets, gravitational microlensing observed by Gaia
is the only known way to precisely determine the low-mass end of 
the stellar mass function in the solar neigbourhood 
(Belokurov\,\&\,Evans 2002), etc. 

But viewed from the basic goal of measuring
star parallaxes and space velocities in our galaxy, such effects are
just noise which may limit the ultimately attainable precision
of the target quantities parallax and proper motion.
Thus, if we were to assign a motto to the present paper, it would
be: ``One person's signal is another person's noise''.  
In the following sections we try to give an assessment of the 
significance of the various known effects that may produce
astrometric fluctuations. In some circumstances they can become 
really disturbing, as we will see. 

Before discussing the above-mentioned effects we present a few general considerations
(Section~2), and in closing we very briefly discuss possible other effects
(Section~7). 
 
\section{General Considerations}

All three categories of effects mentioned in the introduction produce
photometric disturbances as well as astrometric ones. This paper
concentrates on the astrometric effects. But it should be kept in mind
that the quasi-simultaneous photometric measurements of Gaia will in
general give important clues to the underlying cause of any detected
astrometric noise.

Parallax measurements are disturbed by any astrometric noise that can
be interpreted as a Fourier component at 1\,year period (and at the 
correct phase and position angle aspect ratio for a parallax ellipse).
It is not needed that the true astrometric displacement has a 
component at that frequency, due to the convolution of the true 
motion's Fourier spectrum with the strange sensitivity function
generated by Gaia's sky scanning law. Truly 1-year periodic
motions will be the worst case, of course.

Similarly, space velocity measurements are disturbed by anything 
that mimics a mean motion over Gaia's mission duration. In other words,
anything that gives a zero-frequency component after convolution 
with the sensitivity function. Again, truly constant motions will
give the biggest effects.

We stated that the goal of the present paper is to give an `assessment' rather
than a full quantitative analysis. This is for two reasons: Firstly, for some
of the effects a precise quantitative estimation is presently not possible.
Secondly, it would be misleading to simply give a median or rms value. The size
distribution of the angular disturbances does not fall off steeply
at large values, unlike e.g.~a Gaussian: A given linear displacement
of the photocentre in space leads to an angular displacement that is
inversely proportional to distance. Thus, for a roughly homogeneous
distribution of the observed stars in space, the size distribution of the 
angular displacements falls off as their fourth power. Such a 
distribution has 0.15~percent remaining cases (i.e.~$\sim$1.5~million
stars out of Gaia's one billion) beyond 5~times its rms value, while 
a Gaussian distribution has only about 10$^{-7}$ cases 
(i.e.~about 100 stars for Gaia) beyond 5~times its rms value. 
The `tail' of large values is caused by nearby stars.

\section{Stellar Granulation}

Stellar convection zones reaching up to the photosphere cause the
irregular and time-varying surface brightness pattern called granulation.
It produces a random offset of the centre of light of a stellar
disk (which is observed by Gaia) from the centre of mass of the star
(which would define an undisturbed parallax ellipse and galactic space 
velocity). The size of the astrometric effect is easy to quantify, as
shown by Svensson\,\&\,Ludwig (2004). Its time scale is the convection
cell lifetime at the surface. Fig.~1 shows a simulation of the 
astrometric offset, derived from a radiation-hydrodynamical model of 
a red giant star by Svensson\,\&\,Ludwig. 

 \begin{figure}[h]
  \begin{center}
    \leavevmode
\centerline{\psfig{file=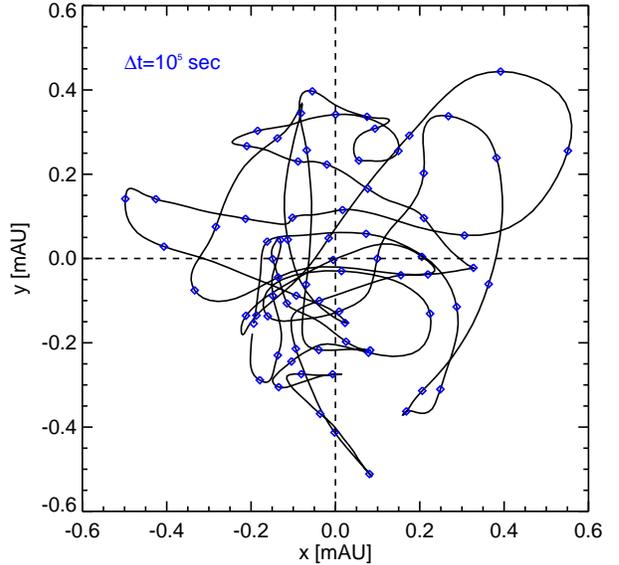,width=8cm,bbllx=40pt,bblly=20pt,bburx=570pt,bbury=530pt,clip=}}
   \end{center}
  \caption{Simulated astrometric displacement of the photocentre of a red giant (see text) from its
  centre of mass, with ticks added to the curve every 10$^5$ seconds 
  (from Svensson\,\&\,Ludwig 2004).}
  \label{fig1}
\end{figure}

The surface of a convective star is fully covered with convection cells.
Each individual cell produces an rms photocentre offset from the star's 
centre of gravity
of $\sqrt{1/6}$ times the radius $R_{\rm star}$ of the star. So, for the average over
$N_{\rm cell}$ statistically independent cells of typical size $r_{\rm cell}$ and brightness
contrast $C$ we have approximately
\begin{eqnarray} 
\sigma_{astrom} & \sim & C {1\over \sqrt{6}} R_{\rm star} N_{\rm cell}^{-1/2} \phantom{blablabla} \\
                & \propto & C {1\over \sqrt{6}} R_{\rm star} (r_{\rm cell}^{-2} R_{\rm star}^2)^{-1/2}\\
		& \propto & C {1\over \sqrt{6}} r_{\rm cell} \phantom{blablablablablabl}
\end{eqnarray}
Therefore the radius of the star cancels out: the astrometric disturbance depends on 
$r_{\rm cell}$ only (the contrast $C$ in the optical range 
always being of the order of unity).
The cell size is proportional to the surface pressure scale height, which in turn is 
inversely proportional to the surface gravity (with a small dependence on the mean
molecular weight). Thus the rms astrometric displacement for any star with granulation
can be expected to be inversely proportional to log~g.
This simple analytical consideration of Svensson\,\&\,Ludwig
is fully confirmed by their numerical simulations which are summarized in Fig.~2.

 \begin{figure}[h]
  \begin{center}
    \leavevmode
%\centerline{\psfig{file=bastian_figure2.eps,width=8cm,angle=90,bbllx=30pt,bblly=30pt,bburx=530pt,bbury=73pt,clip=}}
\centerline{\psfig{file=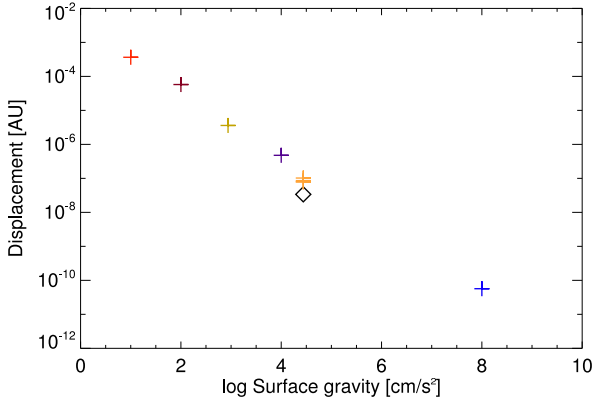,width=5.6cm,angle=90,clip=}}
   \end{center}
  \caption{Root-mean-square astrometric displacements of stars due to granulation as function
  of surface gravity (from Svensson\,\&\,Ludwig 2004). For details see text. 
  The diamond symbol shows the weak effect of
  metallicity ([Fe/H]=-2 instead of zero) for the case of solar-type stars.  }
  \label{fig2}
\end{figure}

That figure clearly shows that the granulation noise is negligible for 
white dwarfs (log\,g\,=\,8), 
main-sequence stars (log\,g\,=\,3...5) and Cepheids 
(log\,g\,=\,2). For a red giant star like that of Fig.~1 
(T$_{\rm eff}$=3676\,K, log\,g=1) 
the effect is of the 
order of 1\,mAU~=~10$^{-3}$ astronomical units,
and its timescale is of the order of a day. This 
is measurable for nearby giants (since 1~mAU translates 
into 10\,microarcsec 
($\mu$as) at 100\,pc), but is still irrelevant. However,
for the parallaxes of Mira stars and cool supergiants (log\,g around zero)
it becomes problematic: Firstly, the rms displacement approaches
a considerable fraction of an astronomical unit. But even worse, 
its typical time scale
gets close to one year. So it may create significant parallax errors.
For space velocity studies it is irrelevant for all types of
stars because even 0.1\,AU/5\,years is only 100\,m/s.

\section{Star Spots}

The astrometric effects of star spots (and other magnetic surface 
features) are hard
to quantify statistically, simply because the statistics of star spots are 
insufficiently known. The most specific statements on the distribution of numbers,
sizes etc. versus star type that we could find in the literature (e.g.~in   
various papers in Strassmeier et al.~2002): 
\begin{itemize}
\item ``... are common among cool stars with outer convection zones.''
\item complaints: all available data are strongly 
biased towards very active stars.
\item discussions about supergiants: Do we see dark spots or bright cells?
\end{itemize}
In the absence of good statistics one can still get a semi-quantitative 
assessment by looking at some representative cases. 

\begin{itemize}
\item For the sun, a spot covering $f$=1~percent of a hemisphere
is unusually large. Such a spot can produce a {\em maximum} astrometric offset
of 0.5$f r_\odot$\,=\,2.5\,10$^{-5}$\,AU; this is completely negligible.
\item For a solar-type star with a huge spot the maximum displacement
may reach up to 0.5$ r_\odot$\,=\,2.5\,10$^{-3}$\,AU. For the sake of 
parallax or space velocity measurements this is still negligible, but it is
clearly measurable with Gaia in some cases: 2.5\,10$^{-3}$\,AU translate into
100\,$\mu$as at 25\,pc. The periodic displacement of the centre of light of 
a rotating spotted star might then be mistaken as the gravitational pull 
of a low-mass companion. 
\item K giants are roughly ten times as big as solar-type stars. Large spots
on their surface will thus be readily detectable astrometrically 
by Gaia. They will still be quite harmless for the
measurement of parallaxes and space motions.
But the danger of confusion with low-mass companions will be more severe.
\item For supergiants and M giants, having radii of the order of 100\,$R_\odot$
(and more), the effect may reach 0.25\,AU (or more). If its time dependence
should happen to mimick a parallax motion, very significant distance errors
may result. The maximum possible effect will rarely occur, but a noticeable extra 
parallax noise can be expected for these stars if they carry large spots (or
large convection cells, see previous section). 0.25\,AU/5\,years translate into
50\,$\mu$as/year or 250\,m/s at 1\,kpc.  
\end{itemize}

In closing this section let us add three remarks: 1) Big star spots
are connected with brightness changes, so the possible confusion with low-mass
companions or parallax motion can be (largely) avoided by 
Gaia's parallel photometry.
2) Plagues tend to be 
larger than spots (at least in the case of the sun), but have much smaller 
contrast; so they should have smaller effects. 
3) The magnetically most active star types known, BY Dra stars and 
RS CVn stars, are small in size and are quickly rotating; so their 
spots are astrometrically irrelevant.

\section{Gravitational Lensing}

\subsection{Basics}

Gravitational lensing occurs when a gravitating body (the ``lens'') 
comes close to the 
line of sight between a light-emitting body (the ``source'') and an 
observer. It generally 
leads to a splitting and focussing of the light path between source and observer. 
The observer then sees two (or more) images of the source, usually at higher total
brightness than without the lens. The brightness increase due to 
gravitational lensing
has been systematically searched for and detected by large 
specialized surveys like OGLE and
MACHO. 
For a point-like lens the (two-dimensional) astrometric 
displacement 
of the centre of light $\vec{\theta}(t)$ at time $t$ is given by
(see e.g.~Walker 1995):
\begin{equation}
\vec{\theta}(t)=\frac{\vec{\theta}_{\rm SL}(t)}
{[\theta_{\rm SL}(t)/\theta_{\rm E}]^2+2}
\end{equation}
where the arrows indicate small tangential vectors on the celestial sphere,
$\vec{\theta}_{\rm SL}(t)$ is the angular separation of the (unlensed) source 
from the lens
at time $t$, and $\theta_{\rm SL}(t)$ its absolute value.
The Einstein radius $\theta_{\rm E}$ depends on the lens mass $M$ by
\begin{equation}
\frac{\theta_E}{\rm 1\,mas}=\left(\frac{M}{0.12 M_\odot}\right)^{1/2}
\left(\frac{\varpi_{\rm L}-\varpi_{\rm S}}{\rm 1\,mas}\right)^{1/2}
\end{equation}
where $\varpi_{\rm S}$ and $\varpi_{\rm L}$ are the parallaxes of the 
source and lens, 
respectively, and `mas' is the abbreviation for milliarcsecond. The 
time dependence of
$\vec{\theta}$ is due to the difference of proper motion and parallactic shift
between source and lens.

Gravitational lensing comes in three observational
``flavours'':
\begin{itemize}
\item Macrolensing: The effect is called macrolensing if the separation 
between the 
split images is large enough that they can be directly resolved, i.e. if it
approaches the level of an arcsec. In practice this happens only if the lens 
is of galaxy-sized mass. It leads to the well-known multiple images of 
quasars behind foreground galaxy clusters, of which Gaia will observe a
considerable number. The opposite case, i.e. when the split 
images are unresolved,
is called microlensing.
\item Strong microlensing: The effect is called strong microlensing when 
$\theta_{\rm SL}$ is of the order of the 
Einstein radius $\theta_{\rm E}$. The displacement of the centre 
of light then is of the order of 
$\theta_{\rm E}$ as well.
\item Weak microlensing: When $\theta_{\rm SL}$ 
is much larger than $\theta_{\rm E}$ the effect is called weak
microlensing. In this case the displacement of the 
centre of light is of the order of
$\theta_{\rm E}(\theta_{\rm E}/\theta_{\rm SL})$.
\end{itemize}

\subsection{Stars}

The subject of astrometric microlensing with Gaia has been thorougly investigated
by Belokurov\,\&\,Evans (2002). From extensive simulations they find that strong 
microlensing will produce very significant and astrophysically extremely useful 
effects, but at the same time they find that it is very rare. 
They define a significance
level by requesting that the maximum astrometric displacement be $7\sigma_{\rm FoV}$,
where $\sigma_{\rm FoV}$ is Gaia's (magnitude-dependent) measuring precision from
a single field-of-view transit of a star, and that this maximum displacement occurs
within Gaia's mission duration of five years.

From this criterion they find that one out of every 40\,000 Gaia stars will be
significantly lensed during the mission (in the microlensing jargon: the optical
depth is 2.5\,10$^{-5}$). In other words: Gaia will observe about 25\,000 
highly significant events, with about 2\,500 of them yielding precise lens masses.
The most important lenses are found to be low-mass stars
within a few hundred pc from the sun. The most important sources are disk and
bulge stars several~kpc away.   

\begin{figure}[t]
  \begin{center}
    \leavevmode
\centerline{\psfig{file=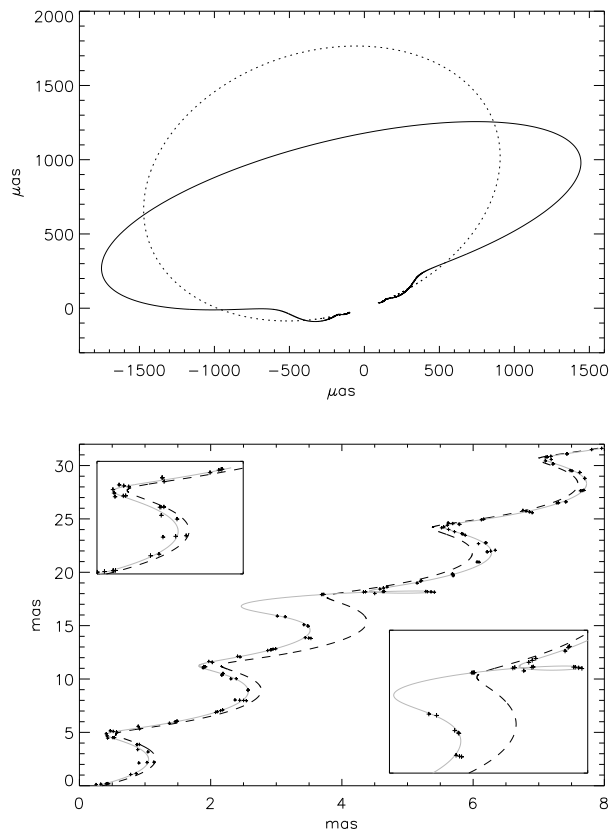,width=8cm,bbllx=70pt,bblly=150pt,bburx=465pt,bbury=695pt,clip=}}
   \end{center}
  \caption{Upper panel: astrometric shift of a microlensing event by a lens of
  0.5 solar masses at 150\,pc and a source at 1.5\,kpc, with 70\,km/s transverse
  velocity, as seen by a barycentric observer (dotted line) and Gaia (solid line).
  The difference is due to the differential parallactic shift. Lower panel: The same,
  but showing the actual path of the source's centre of light on the sky, including
  proper motion and parallax (grey line). The dashed line indicates the unlensed
  path of the source. The tiny crosses show simulated Gaia measurements, with 
  $\sigma_{\rm FoV}$=300\,$\mu$as corresponding to a 17th-magnitude star. The 
  insets show enlarged views of the beginning and the midpoint of the event  
  (from Belokurov\,\&\,Evans 2002).
  }
  \label{fig3}
\end{figure}

To give just a few specific numbers: The maximum positional shift during a 
microlensing event occurs when  $\theta_{\rm SL}(t) = 1.41 \theta_{\rm E}$.
Its absolute value then is $0.35\theta_{\rm E}$, which for a distant source 
amounts to 
about 1\,100~$\mu$as for an 0.12\,solar-mass lens at 100\,pc, and to 
350~$\mu$as for an 0.12\,solar-mass lens at 1\,kpc. 
The astrometric appearance of a microlensing event is as follows 
(Fig.~\ref{fig3}):
Over a few years
the photocentre of the source moves away from the unlensed position with
increasing speed, then within a few months it performs a quick ``swing'' to the
opposite side, and finally it moves back towards the unlensed position with
quickly decreasing speed.

Belokurov\,\&\,Evans find that strong microlensing, being so rare, is of
negligible effect for the overall error budget of Gaia. In the few cases
where its size is significant it will rarely be mixed up with a proper 
motion or parallax effect\footnote{Less significant events and
incomplete events (i.e. events with the maximum
displacement outside Gaia's mission lifetime) can be mistaken as binary
motion, however.}.

Weak microlensing, on the other hand, is ubiquitous, but small. The typical
angular displacement of a Gaia star by weak microlensing is of the order
of 1\,$\mu$as, with a secular change of the order of 0.01--0.1\,$\mu$as/year.

\subsection{Quasars}

Quasars were once thought to be perfect inertial reference points, with 
zero parallaxes and proper motions. But in fact
they are not. A proper-motion scatter somewhere in the range 10--100\,$\mu$as/year must
be expected for them, from a variety of causes (quite a number of publications
relevant for high-precision astrometry of quasars are cited and summarized in
the Gaia Concept and Technology Study Report (ESA 2000), p. 109-119):
\begin{itemize}
\item random proper motions due to relativistic jets, up to 500\,$\mu$as/year, but mostly 
small (extra noise; no bias)
\item systematic proper motions of 3$\mu$as/year on average due to the galactocentric
acceleration of the sun (bias, but no problem; will be modelled) 
\item image centroiding problems due to underlying galaxy (extra noise; no bias)
\item image centroiding problems due to non-stellar spectra 
(imperfect instrumental chromaticity correction, extra noise; no bias)
\item random proper motions due to weak microlensing by galactic stars 
(typically 0.01--0.1$\mu$as/year, negligible; no bias) 
\end{itemize}
The above items are present for all quasars. Those quasars which are
macrolensed by intervening
galaxies (about one percent of all, i.e. several thousand out of Gaia's 
total of 500\,000 
or so) suffer additional astrometric noise:
\begin{itemize}
\item image centroiding problems due to non-pointlike appearance
(extra noise; no bias)
\item time variability of the macrolensing due to relative tangential motion
of quasar and lens, leading to changes of the relative brightness of the 
quasar images (random proper motions, often one or a few $\mu$as/year, up to
dozens of $\mu$as/year; no bias)
\item time variability of the macrolensing due to strong microlensing by individual 
stars in the lens galaxy, again leading to changes of the relative brightness of the 
quasar images (character and size of the effect as in the previous item)
\end{itemize} 

Quasar astrometry will be the central tool to make Gaia's rigid astrometric sphere kinematically non-rotating,
by using the quasars as a grid of sources with ``zero'' proper motion. 
All the above effects (except the galactocentric acceleration of the sun,
see Bastian 1995) will add to the measurement noise in quasar proper motions, making
the quasars less favourable as might naively be expected. Fortunately, the extra
noise is at most as large as the measurement noise (the latter being of the order
of 100--200\,$\mu$as/year for quasars of 18th to 20th magnitude), and will 
cause no bias.

Concerning parallaxes, Gaia intends to produce absolute parallaxes without external
reference. However, the quasars will provide a useful data set to check the 
success in this respect. Again, the above-listed effects will cause extra noise,
but essentially no bias. It should be mentioned that Sazhin et al. (2001) 
recently discovered that on parallaxes the weak microlensing effect
does not create just an extra noise, but also a bias. 
Due to this effect all quasars are expected
to exhibit very small (typically a few nano-arcsec) negative parallaxes. 
Sazhin et al.~demonstrate that
in exceptional cases (perhaps one percent of all quasars, maybe less) 
this negative pseudo-parallax may reach 1\,$\mu$as. However, its typical value will
be irrelevant for Gaia. The average value can be roughly estimated to be 
between 10~and 30\,nanoarcsec. With 500\,000 quasar parallaxes individually 
measured to 100\,$\mu$as precision (optimistic for Gaia), the random error of
their weighted average will be of the order of 150\,nanoarcsec. So the bias due
to weak microlensing is negligible\footnote{in the oral presentation of the present 
paper during the conference we claimed it to be relevant; but this was due to
a misunderstanding of some statements in Sazhin et al.~2001.}.

\section{Stellar Multiplicity}

Orbital motion of double and multiple stars causes measurable deviations of the 
observed stellar positions from a pure proper motion and parallax model which,
if undetected or at least unmodelled, will create errors in the measured parallaxes
and space velocities. As we can safely assume binary orbits to be randomly 
oriented in space and randomly phased in time, we need not fear any biases
to be introduced in this way. If the binary nature of the motion is 
detected and correctly modelled it will not even create extra noise (but will
still weaken the determination of parallaxes and proper motions due to the 
increased number of unknowns to be solved for).

\subsection{Space Velocities}

The effects of stellar multiplicity are being carefully studied by the 
Gaia Double and Multiple Stars Working Group, see the paper by S{\"o}derhjelm 
in this volume and references therein. Extensive simulations of
Gaia measurements on double stars, and of binary detection and
orbit modelling performed on those measurements, have been carried out. They
resulted in the general picture summarized in Figs.\,4 and\,5.

Fig.\,4 shows the statistics of errors in the measured tangential velocities (of the 
simulated stars) created by undetected binarity as function of orbital period, for
bright stars only (magnitude~10 to~12.5). 
The median error due to measurement noise is indicated as a thin horizontal line.
At very long periods Gaia cannot detect the orbital motion 
(but generally sees the 
two binary components as separate light sources). 
The extra space velocity errors are not important
because the orbital velocities are very small. At successively shorter periods
(down to about one hundred years) the orbital motion increases on a Keplerian
slope ($v \propto P^{-1/3}$), but still cannot be distinguished from
rectilinear motion in space. The typical error now is much larger than the 
measurement noise. At periods shorter than a few dozen years, Gaia 
quickly starts to recognize the 
orbital acceleration. This creates the prominent gap in the cloud of points in Fig.~4.
For periods around the mission lifetime of Gaia there are practically no
undetected binaries. 

At still shorter periods the detection efficiency becomes worse again because the
astrometric size of the orbits shrinks below the detectability limit. The space
velocity errors introduced by the binarity are nevertheless small because Gaia
sees the motion averaged over many orbital revolutions.

Fig.~5 shows an analogous plot for faint stars (magnitude~17.5 to~20). The same
mechanisms as in Fig.~4 are in effect, but the median measurement error
(thin horizontal line) is about two powers of ten higher than for the bright 
stars. This has two independent reasons: Firstly, the (angular) measurements of
the faint stars are about an order of magnitude less precise. Secondly, the 
faint stars are typically about an order of magnitude farther away from
the sun. As a consequence the binarity-induced errors are almost negligible
compared to the measurement noise.    

 \begin{figure}[h]
  \begin{center}
    \leavevmode
\centerline{\psfig{file=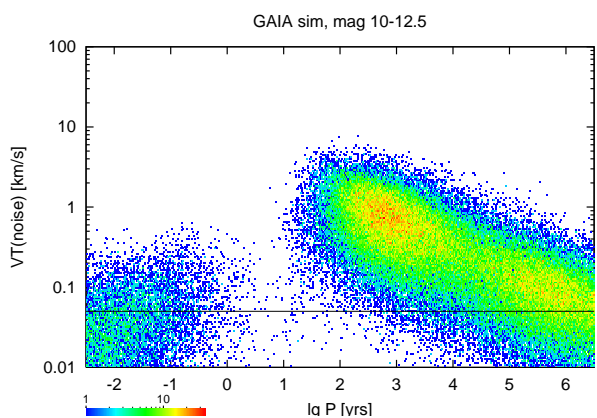,width=5.6cm,bbllx=200pt,bblly=50pt,bburx=554pt,bbury=554pt,angle=270,clip=}}
   \end{center}
  \caption{Space velocity errors induced by undetected binarity, for bright stars, 
  plotted versus binary
  orbital period. The dense parts of the cloud of points are encoded in
  gray scale. 
  Simulations by S. S{\"o}derhjelm, copied from ESA (2000), Fig.\,1.27.}
  \label{fig4}
\end{figure}

 \begin{figure}[h]
  \begin{center}
    \leavevmode
\centerline{\psfig{file=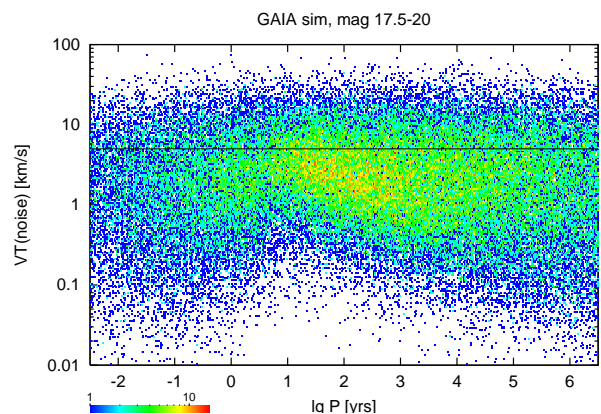,width=5.6cm,bbllx=200pt,bblly=50pt,bburx=554pt,bbury=554pt,angle=270,clip=}}
   \end{center}
  \caption{Same as Fig.\,4, but for faint stars. 
  Simulations by S. S{\"o}derhjelm, copied from ESA (2000), Fig.\,1.27.}
  \label{fig5}
\end{figure}

We note that Figs.\,4 and~5 are based on a now obsolete Gaia hardware configuration and on
a slightly too optimistic binary detection efficiency estimate. An updated study by 
S. S{\"o}derhjelm (private communication) yielded differences 
in the details, but no changes in
the basic aspects and conclusions. Updated graphics may possibly be included in the 
paper by S{\"o}derhjelm in this volume.

\subsection{Parallaxes}

Binary motion will likewise produce extra noise in the determination of parallaxes.
This is relevant for fairly short orbital periods only. Recent simulations by
S. S{\"o}derhjelm (unpublished) show that 
for moderately bright stars (brighter than magnitude~15.5) the typical effect of 
undetected binarity is of the order of one or two dozen~$\mu$as, 
i.e.~comparable to the 
measurement noise. For periods close to one year ($\pm$10\,percent) 
some parts of the orbital motion become
indistiguishable from parallactic shifts. Very big parallax errors 
(up to hundreds of $\mu$as) occur in such
cases. The big errors occur even for {\em detected} binaries of about 1~year
period, because in these cases the derived orbital parameters become strongly 
correlated with parallax.

\section{Other Effects}

There are a number of other physical effects that can create
fluctuations of star positions. In this section we very briefly list 
them and comment on them.    

\begin{itemize}
\item Asymmetric outbursts, gas jets etc.: Rare, except for nearby quasars,
where motions of a few hundred $\mu$as in a few months may occur.
\item Interstellar and interplanetary scintillation: Important in radio
astronomy but negligible in the optical (inversely proportional to the 
frequency of the observed radiation).
\item Primordial gravitational radiation: Predicted to have been produced 
by cosmic inflation, but at amplitudes many orders of magnitude 
too small to be observed by Gaia.
\item Local gravitational radiation, e.g. from neutron star or black-hole
binaries: Highly improbable; would require such a binary to move extremely
close to the line of sight of a background source.
\item Unexpected effects? Going to entirely new realms of precision and/or
numbers of measurements always bears the possibility to discover completely
unexpected phenomena. If this should happen for Gaia, it might --- depending
on the size of the unexpected effect ---  
either be welcomed as a great scientific achievement, or 
regarded as a nuisance for
measuring stars and the Galaxy, or even both.  
\end{itemize}

\section{Conclusion} 

In summary, there are a number of astronomical phenomena that produce
genuine fluctuations of star positions comparable to or larger
than Gaia's measurement noise, at least for parts of Gaia's target
celestial objects.

\begin{itemize}
\item Binarity is a major problem for the astrometry of bright stars.
It will significantly impair the kinematics of stellar aggregates
with small velocity dispersion (small star clusters, associations,
star-forming regions). It will also produce a small proportion of
grossly wrong parallaxes.
\item Stellar granulation is no problem except for a small number
of cool supergiants. It will be a serious problem for e.g. Mira star
parallaxes. 
\item Star spots are hard to assess quantitatively. They may be
problematic for supergiants and very cool giants.
\item Gravitational microlensing will not be a problem, although
thousands of highly significant microlensing events will be detected
by Gaia.  
\item Gravitational macrolensing of quasars will add some noise
to their (otherwise ``zero'') proper motions and parallaxes.
\end{itemize}

\section*{Acknowledgements}

We would like to thank H. Ludwig for providing a draft paper before
its publication, S. Jordan for critically reading the manuscript,
and M. Perryman, W. Evans, S. S{\"o}derhjelm,
V. Belokurov and L. Lindegren for data and graphics.

\end{document}